\documentclass[aps,preprint,amssymb,showpacs]{revtex4} 

\begin{document}

\title{Fluctuating Dimension in a Discrete Model\\
for Quantum Gravity Based on the Spectral Principle}
\author{Luiz C. de Albuquerque$^1$, Jorge L. deLyra$^2$, and 
Paulo Teotonio-Sobrinho$^2$}
\affiliation{(1) Faculdade de Tecnologia de S\~ao Paulo - DEG - CEETEPS - 
UNESP,
Pra\c{c}a Fernando Prestes, 30, 01124-060 S\~ao Paulo, SP, Brazil}
\affiliation{(2) Universidade de S\~ao Paulo, Instituto de F\'{\i}sica - DFMA,
Caixa Postal 66318, 05315-970, S\~ao Paulo, SP, Brazil}

\begin{abstract}

The spectral principle of Connes and Chamseddine is used as a 
starting point to define a discrete model for Euclidean quantum
gravity. Instead of summing over ordinary geometries, we consider
the sum over generalized geometries where topology, metric and
dimension can fluctuate. The model describes the geometry of
spaces with a countable number $n$ of points, and is related to
the Gaussian unitary ensemble of Hermitian matrices. We show that
this simple model has two  phases. The expectation value 
$\langle n \rangle$,
the average number of points in the universe,  is finite in one
phase and diverges in the other. We compute the critical 
point as well as the critical exponent of $\langle n\rangle$. Moreover, the 
space-time dimension $\delta$ is a dynamical observable in our
model, and plays the role of an order parameter. The computation
of $\langle\delta\rangle$ is discussed and an upper bound is found, 
$\langle\delta\rangle\,<\,2$.

\end{abstract}


\maketitle


The quest for a quantum theory of gravity is one of the major
goals in contemporary research. Any attempt in this direction
involves the understanding of the space-time structure at 
very short distance. It is generally
believed that at the Planck scale space-time may not be described
by a manifold \cite{Kempf}. The conventional geometrical setting
of general relativity seems to be inadequate to describe the
non-manifold micro-structures of space-time. Since the manifold
structure has to appear at some macroscopic  limit, it is natural to 
expect that one needs a generalization of ordinary geometry, such as
noncommutative geometry (NCG), rather than a completely new formalism. 

The basic idea coming from NCG \cite{NCG} is that one can describe
a Riemannian manifold $(M,g_{\mu \nu})$ in a purely algebraic way.
There is no loss of information if, instead of the data $(M,g_{\mu
\nu})$, one is given a triple $({\cal A},{\cal H}, D)$, where
${\cal A}$ is the C*-algebra 
$C^0(M)$
of smooth functions on $M$, 
${\cal H}$ is the Hilbert space of $L^2$-spinors
on $M$, and ${\cal D}$ is the Dirac operator acting on ${\cal
H}$. From the Gelfand-Naimark theorem it is known that
the topological space $M$ can be reconstructed from 
the set $\hat{\cal A}$ of
irreducible representations of $C^0 (M)$.
Metric is also encoded, and the geodesic distance can
be computed from  ${\cal D}$.  Here 
we will consider only commutative spectral triples - this
is enough to go much beyond ordinary geometry. In particular one 
can treat all Hausdorff topological spaces in this way. Given a pair
$(M,g_{\mu \nu})$, one can promptly construct the corresponding
triple $(C^0 (M),L^2(M),{\cal D})$. However, not all commutative
spectral triples, or generalized geometries, come from a pair 
$(M,g_{\mu \nu})$. Nevertheless one can always associate a 
Hausdorff space $M=\hat {\cal A}$ to a commutative spectral triple, 
where $\hat {\cal A}$ denotes the set of irreducible representations
of ${\cal A}$. However, the space $M$ may not be a manifold. 
Once we
trade the original Riemannian geometry for its corresponding
commutative triple we need a
replacement for the Einstein-Hilbert action $S_{EH}$.
The so-called spectral action 
of Chamseddine and Connes \cite{CC} is one possible candidate. 
It depends only on the eigenvalues of ${\cal D}$ and contains 
$S_{EH}$ as a dominant term. In this paper however we shall 
use another spectral action.

The spectral action can be 
written for any triple, regardless of whether it comes from a manifold 
$(M,g_{\mu \nu})$ or not. In the spectral geometry approach  it is 
conceivable to write the partition function
\begin{equation}\label{1}
Z=\sum_{x\in {\cal X}}e^{-S[x]},
\end{equation}
were the \lq\lq sum'' is over the set ${\cal X}$ of all possible
commutative spectral triples and $S$ depends on the spectrum of
${\cal D}$. It includes all Hausdorff spaces
and therefore all manifolds of all dimensions.

The framework of spectral NCG is appropriate
to explore some difficult questions in quantum gravity. 
For example,  Rovelli used a simple dynamical model based on
a finite dimensional spectral triple to construct a quantum theory
with a quantized physical distance between points \cite{Rovelli}. 
In our model we look at the dimension as a dynamical quantity and study its
expectation value.

The exact computation of (1) is a formidable task, not yet
accomplished. However, the algebraic approach suggests
ways  of defining discrete approximations to the full theory.
For instance one may replace the algebra ${\cal A}$ by a finite
dimensional algebra $A_n$. In this approach to discretization
there is no need to introduce a lattice or simplicial
decomposition of the underlying space. The approximation of ${\cal
A}$ by a finite algebra works even if the spectral triple does not
come from a manifold. Thus, it gives us a generalization of
ordinary discretizations \cite{discrete}.

In this paper we  discretize  (\ref{1}) by sampling the
set ${\cal X}$ with finite commutative spectral triples.
We will think of it as a useful toy model, which we believe 
captures some of the main features of the full one, Eq. (1). 

The key
role played by the eigenvalues of the Dirac (or Laplace) operator
in the spectral action approach was emphasized in \cite{Landi}. In
our model they are also the natural dynamical variables due to the
connection with random matrix theory (RMT). An important
ingredient of the model is that the number of points can
fluctuate.  Moreover, in our simple model the space-time dimension
is a dynamical  observable and its expectation value can be  computable 
from first principles.

Let us describe the ensemble $X\subset \cal{X}$ of geometries we 
will consider.
A point of $x\in X$ is a commutative spectral triple $x=({\cal
A},{\cal H}, D )$ where the commutative \mbox{C*-algebra} ${\cal
A}$ has a countable spectrum ${\hat {\cal A}}$. We divide $X$ into
subspaces $X_n$ consisting of triples $({\cal A}_n,{\cal H}_n, D
)$ such that ${\hat {\cal A}}_n$ has a fixed number $n$ of points.
From the Gelfand-Naimark theorem it follows that elements of
${\cal A}_n$ are the (possibly infinite) sequences
$a=(a_1,a_2,...,a_n)$, $a_j\in{\mathbb{C}}$. The Hilbert space
${\cal H}_n$ is given by vectors $v=(v_1,...,v_n)$ with norm
$||v||^2\equiv\sum_{i=1}^n v_i^2 <\infty$. The elements of ${\cal
A}$ are represented by diagonal matrices $ \hat
a=\mbox{diag}(a_1,...,a_n)$ acting on ${\cal H}_n$. Finally, the
operator $D$ is a $n\times n$ self-adjoint matrix. We will sample
the space $X$ by $X_1,X_2,...,X_N$ and eventually take the limit
$N\rightarrow \infty $.

Let $L$ be a length scale
such that the operator ${\cal D}$
given by ${\cal D}=D/L$ will be the analogue of the Dirac
operator. The Chamseddine--Connes action
depends on a cutoff function of the eigenvalues of $D/L$.
The cutoff function is zero for eigenvalues of $D$  greater than 
$L$ and one otherwise \cite{CC,Landi}.
In other words, the Boltzmann weight in Eq.(\ref{1}) 
would be one outside a compact region in the eigenvalue space, 
leading to a divergent partition function 
(see Eq. (\ref{21}). Let us consider a quadratic
action instead:
\begin{equation}\label{4}
S[x]={\rm Tr}\left(\frac{{\cal D}}{\Lambda}\right)^2\equiv 
\beta ~{\rm Tr}(D^2),
\end{equation}
where $\Lambda $ is the inverse of Planck's length $l_p$, and
$\beta = (l_p/L)^2$. Finally, we define the partition function
$Z_N(\beta )= \sum _{n=0}^{N}z_n(\beta )$
where
\begin{equation}\label{5}
z_n(\beta)=
\int [dD] 
e^{-\beta{\rm Tr}(D^2)}
\end{equation}
is the partition function restricted to $X_n$,
in other words,
an integral over all independent matrix elements
$D_{ij}$, where  $[dD]$ is the usual measure for $n\times n$ Hermitian
matrices \cite{Meh}. The partition function $z_n(\beta)$ 
defines the one-matrix Gaussian Unitary Ensemble (GUE)
\cite{Meh}. A straightforward computation gives
$
z_n(\beta)= 2^{\frac{n}{2}}(\frac{\pi }{2\beta})^{\frac{n^2}{2}}.
$

The expectation values of an observable  
${\cal O}(D_{ij})$ restricted to $X_n$ and for the entire ensemble
are 
\begin{eqnarray} \label{pts1}
\langle{\cal O}\rangle_{n,\beta}&\equiv& 
{\int [dD] {\cal O}e^{-\beta
{\rm Tr}(D^2)}}/{z_n(\beta)}~~\mbox{and}\\
\label{pts2}
\langle{\cal O}\rangle(\beta )&\equiv&\sum_{n=1}^{N}
P(n,\beta )\langle{\cal O}\rangle_{n,\beta},
\end{eqnarray}
respectively, where the function $P(n,\beta )=\frac{z_n(\beta)}{\sum _n
z_{n}(\beta)}$ is interpreted as the probability of having a
universe with $n$ points. The simplest observable in our model is
$n$, the number of points in $\hat {\cal A}$. By definition, $n$
is constant in $X_n$, therefore $\langle n\rangle_{n,\beta}=n$. 
Thus we get
\begin{equation}\label{11}
\langle n\rangle(\beta)=\frac{\sum _n n 2^{\frac{n}{2}}
(\frac{\pi}{2\beta})^{\frac{n^2}{2}}} 
{\sum _n 2^{\frac{n}{2}}(\frac{\pi}{2\beta})^{\frac{n^2}{2}}}.
\end{equation}
The mean $\langle n\rangle$ (\lq\lq average number of points in the 
universe'')
is not a continuous function of $\beta$ at $\beta_c=\pi/2 $, signaling the
onset of a phase transition.
Besides straitforward numerical calculation,
there are other ways to show that the
sum (\ref{11}) converges for $\beta>\beta_c$  and diverges for
$\beta<\beta_c$. Consider, for instance, the approximation of
$Z_N(\beta)$ in the limit $N\to\infty$ by
the first term of a Euler-Maclaurin expansion,
\begin{equation}\label{12}
Z(\beta)=\lim_{N\to\infty}\,\sum_{n=0}^{N} z_n(\beta)
=\int_0^\infty dx \,{\rm e}^{-\frac{1}{2}a_{\beta}\,x^2+bx}+
{\cal R}(\beta),
\end{equation}
where $a_{\beta}=\ln(\beta/\beta_c)$, $b=\ln 2/2$. It is easily
seen that
\begin{equation}\label{13}
\langle n\rangle(\beta) =\frac{\partial}{\partial b}\,\ln\, Z(\beta),
\end{equation}
neglecting the remainder ${\cal R}$ in
(\ref{12}) \cite{footnote}. Eq.(\ref{12}) suggests a nice interpretation.
For $\beta>\beta_c$ ($\beta<\beta_c$) the quadratic term is
positive (negative) and the integral converges (diverges). The
phase transition at $\beta=\beta_c$ is triggered  by the change of
signal of a bilinear term in the \lq\lq fields''. The integral in 
(\ref{12}) can be solved in the region $\beta>\beta_c$. 
After expanding $a_{\beta}$ around the critical point, 
$a_{\beta}=(\beta-\beta_c)/\beta_c+...$, 
we compute $\langle n\rangle$ by means of (\ref{13}). The result is
$
\langle n\rangle(\beta)\sim\Bigl(\frac{\beta-\beta_c}{\beta_c}\Bigr)^{-1}
\equiv \tau^{-1}.
$
 Hence, for  $\beta>\beta_c$ the system is in a \lq\lq
finite" phase, characterized by a finite value of $\langle n\rangle$. As
$\beta\to\beta_c^{+}$  $\langle n\rangle$ diverges with a (mean field) 
critical
exponent $\nu=1$. The r.m.s. deviation of $\langle n\rangle$ may be 
computed, with the result 
$\Delta n\equiv\sqrt{\langle n^2\rangle-\langle n\rangle^2}\sim
\tau^{-1/2}$. However, the relative width of the distribution,
$\Delta n/\langle n\rangle$, decreases like $\tau^{1/2}$. 

For $\beta<\beta_c$ the relevant universes have $\langle n\rangle=\infty$ 
and  $\Delta n/\langle n\rangle=0$. For
a $\infty$-dimensional $D$ one can define the dimension $\delta $
of the space $\hat {\cal A}$ from the eigenvalues of $D$. Let
$\{\mu_0(D),\mu_1(D),... \}$ be the modules of the eigenvalues
(i.e. the singular values) of $D$ organized in an increasing
order. By the Weyl formula \cite{NCG}, the dimension $\delta $  is
related to the asymptotic behavior of the eigenvalues for large
$k$: $\mu_k(D)\approx k^{\frac{1}{\delta }}$. By definition
$\delta =0$ for finite dimensional spectral triples. We can argue
that $\langle\delta\rangle$ is of the form
\begin{equation} \label{19}
\langle\delta\rangle(\beta)=\left\{
          \begin{array}{ll}
          f(\beta) & \mbox{if $\beta<\beta_c$}, \\
          0 & \mbox{if $\beta>\beta_c$}.
          \end{array}
        \right.
\end{equation}
This follows from the fact that for $\beta >\beta _c$ the
probability $P(n,\beta )$ is localized  around some finite $n$.
Hence $\langle\delta\rangle$ works as an order parameter. The value
$\beta_c=\pi/2$ separates $\langle\delta\rangle=0$ from the rest.

In order to study the  dimension we need to consider the
spectral $\zeta$-function

\begin{equation}\label{20}
\zeta(z)=\lim_{n\to\infty}\,
\sum_{k=0}^{n} \mu_k^{-z}={\rm Tr}\,(|D|^{-z}),
\end{equation}
where $D$ is an $\infty$-dimensional matrix ($\mu_0>0$). 
The relation between the dimension and $\zeta(z)$ has been 
discussed in \cite{Connes}. For large
enough values of $\alpha={\rm Re}(z)$, 
${\rm Tr}(|D|^{-z})$ is well defined. One says that $D$
has dimension spectrum $Sd$ if a discrete subset 
$Sd=\{s_1,s_2,...\}\subset{\mathbb{C}}$ exists, such that
$\zeta(z)$ can be holomorphically extended to ${\mathbb{C}}/Sd$.
This definition is consistent with the Weyl formula.
The set $Sd$ has more than a single point when for example
the geometry is the union of pieces of different dimensions
\cite{Connes}. In what follows we will look at an upper bound for
the dimension: It may happens that ${\rm Tr}(|D|^{-\alpha})=0$
for large enough $\alpha$, whereas
for small values of $\alpha$,  ${\rm Tr}(|D|^{-\alpha})= \infty$. 
Eventually, there is a value of
$\alpha$ (say, $\alpha_c$) for which ${\rm
Tr}(|D|^{-\alpha_c})$ is finite and non-zero. 
The upper bound for the  dimension will be $\delta=\alpha_c$.

In order to estimate $\langle\delta\rangle$ by means of (\ref{20}), 
we rewrite (\ref{5}) and
(\ref{pts1}) as integrals over the eigenvalues $\lambda_k$ of $D$.
The procedure is well-known \cite{Meh}, and leads to
($C_n\equiv \pi^{\frac{n(n-1)}{2}}/\prod_{k=1}^n k!$)
\begin{equation}\label{21}
z_n(\beta)=C_n
\int_{-\infty}^\infty [d^n\lambda]\Delta^2(\lambda_k)
e^{-\beta\sum_{i=1}^n \lambda_i^2} \equiv 
C_n\Psi_{n,\beta}~,
\end{equation}

\begin{widetext}
\begin{equation}
\langle{\cal O}(\lambda_i)\rangle_{n,\beta} =\int_{-\infty}^\infty
[d^n\lambda]{\cal O}(\lambda_i)\left\{\frac{2^{\frac{n(n-1)}{2}}
\beta^{\frac{n^2}{2}}}
{\pi^{\frac{n}{2}}\prod_{k=1}^n k!} \Delta^2(\lambda_k)
e^{-\beta\sum_{i=1}^n \lambda_i^2}\right\}\equiv
\int_{-\infty}^\infty
 [d^n\lambda]{\cal O}(\lambda_i)
{\cal P}_{n,\beta}(\lambda_k)~,\label{23}
\end{equation}
\end{widetext}
where $\Delta(\lambda_k)=\prod_{i<j}(\lambda_j-\lambda_i)$ is the
Vandermonde determinant (Jastrow factor), and $[d^n\lambda]\equiv
\prod_{i=1}^n d\lambda_i$.

In RMT, $\Psi_{n,\beta}(\gamma)$ is interpreted as the positional
partition function of an ensemble of equal charged particles (with
positions given by $\lambda_i$) in two dimensions, moving along an
infinite line, in thermodynamic equilibrium at temperature
$\gamma$ - the so-called \lq\lq Dyson gas'' \cite{Dy}. 
Then, ${\cal P}_{n,\beta}(\lambda_1,...,\lambda_n)$
defined in (\ref{23}) is the probability of finding one particle
at $\lambda_1$, one at $\lambda_2$, etc.
The value of $\Psi_{n,\beta}(\gamma)$ is known from the Selberg's
integral. 

In the region $\beta\leq\beta_c$ the partition function $Z(\beta)$
is dominated by $\infty-$dimensional matrices. Thus, one may try
to select the $\infty-$dimensional matrices out of the whole
ensemble, and then compute the mean of the $\zeta$-functions
following (\ref{20}). However, from the standpoint of our
statistical approach this procedure does not seems natural since
the sum over $n$ is a key ingredient in the whole construction.
Hence, we look for a quantity related to the $\zeta$-function that
captures the statistical nature of our model. Let us compute the
mean value 
\begin{equation}\label{25}
\langle{\rm
Tr}_\kappa\,|D|^{-\alpha}\rangle_{n,\beta}\,=
\,\left\langle\sum_{k=1}^n\,|\lambda_k|^{-\alpha}\,
\theta(|\lambda_k|-\kappa)\right\rangle_{n,\beta}.
\end{equation}
Bearing in mind (\ref{20}), we want to study $\langle{\rm
Tr}_\kappa\,|D|^{-\alpha}\rangle_{n,\beta}$ only in the region of
eigenvalues where it is a decreasing function of $\alpha$. Hence,
we introduced an \lq\lq infrared'' cutoff $\kappa$, since there
is a non-zero probability to find a configuration in the volume
$\sum|\lambda_k|\leq \kappa$ around the origin. Using the
symmetry of the integrand in (\ref{25}) under permutations of the
position indexes, we arrive at
\begin{equation}\label{26}
\langle{\rm
Tr}_\kappa\,|D|^{-\alpha}\rangle_{n,\beta}\,=\,\sqrt{\beta}\int_{{\rm
out},\kappa}\,d\lambda_n\,|\lambda_n|^{-\alpha}\,
\sigma_n(\sqrt{\beta}\lambda_n),
\end{equation}
where we have used the definition of the spectral density
$\sigma_n(\lambda)=\left\langle\sum
\delta(\lambda-\lambda_k)\right\rangle_n$. Besides, $\int_{{\rm
out},\kappa}\equiv\int_{-\infty}^{-\kappa}+
\int_\kappa^{\infty}$. 
The computation of (\ref{25}) was reduced to a one-particle
problem. It is known that $\sigma_n(\lambda)=\sum_{k}
\phi^2_k(\lambda)$, $\phi_k(x)$ being the Weber-Hermite
functions  \cite{Meh}. In the large 
$n$-limit, $\sigma_n(\lambda)$ converges
to a nonrandom function, Wigner's \lq\lq semi-circle law'',
\begin{equation}\label{27}
\sigma_n(\sqrt{\beta}\lambda)=\frac{\sqrt{2n}}{\pi}\,
\sqrt{1-\frac{\beta}{2n}\lambda^2},
\end{equation} 
for $|\lambda|<{\cal R}_{n,\beta}\equiv\sqrt{2n/\beta}$, 
and zero otherwise.
The average eigenvalue density for Gaussian ensembles of Hermitian
matrices with  different values of $n$ differ only by a change of
scale, for large enough $n$ (since ${\cal
R}_{n,\beta}\sim\sqrt{n}$). We propose the following quantity in
order to extract information on the $\zeta$-function (\ref{20}):
\begin{equation}\label{28} 
\langle{\rm Tr}_{\kappa_{n,\beta}}
|D|^{-\alpha}\rangle(\beta)=\lim_{N\to\infty}
\sum_{n=1}^{N} P(n,\beta)\langle{\rm
Tr}_{\kappa_{n,\beta}}|D|^{-\alpha}\rangle_{n,\beta}~ .
\end{equation}
Hence, we are sampling the partial traces of the ensemble of
$\infty-$dimensional Hermitian matrices by the total trace of
finite dimensional matrices in the complete Gaussian ensemble of
Hermitian matrices. Now, we study the asymptotic properties of $\langle
{\rm Tr}_{\kappa_{n,\beta}}\,|D|^{-\alpha}\rangle_{n,\beta}$ as a
function of $\alpha$ and $n$ for $n\sim N$ large. In this case,
one may use the semi-circle law in (\ref{26}),
\begin{equation}\label{29}
\langle{\rm Tr}_{\kappa_{n,\beta}} |D|^{-\alpha}\rangle_{n,\beta}
\approx \frac{2}{\pi}\sqrt{2n\beta}
\int_{\kappa_{n,\beta}}^{{\cal R}_{n,\beta}}
d\lambda_n |\lambda_n|^{-\alpha} \sigma_n(\sqrt{\beta}\lambda_n).
\end{equation}
In (\ref{29}) we are neglecting the contribution from the \lq\lq
tail'' of $\sigma_n(\sqrt{\beta}\lambda)$, outside of the
semi-circle radius ${\cal R}_{n,\beta}$. There
is an exponential decrease in the tail, so that the total number
of particles (or eigenvalues) in this region is of order 2
\cite{Bronk}. This finite-size correction is not relevant to our
asymptotic analysis. Besides, the cutoff $\kappa$ is a function
of $n$ and $\beta$:
$\kappa_{n,\beta}=\epsilon\sqrt{2n/\beta}$. The mean
particle (or level) spacing is
$\overline{s}(\lambda,\beta)\sim\sigma_n^{-1}(\lambda,\beta)$, so
that the mean spacing is almost uniform near the origin. Near the
endpoints (edges) the spacing is highly non-uniform. This is the
asymptotic region we are interested in. Our procedure is to select
a slice of size $\Delta_{n,\beta,\kappa}={\cal R}_{n,\beta}-
\kappa_{n,\beta}=(1-\epsilon){\cal R}_{n,\beta}$ near the edge,
so that the relative size of the slice,
$\Delta_{n,\beta,\kappa}/{\cal R}_{n,\beta}$, does not depend on
$n$ and $\beta$. This choice ensures that we are treating matrices
of different sizes $n$ on the same footing. From (\ref{29}) and
(\ref{26}) we obtain, after some manipulations,
\begin{equation}\label{30}
\langle{\rm Tr}_{\kappa_{n,\beta}}\,|D|^{-\alpha}\rangle_{n,\beta}\,
\approx\,
\frac{2}{\pi}\,(2n)^{1-\frac{\alpha}{2}}\,\beta^{\frac{\alpha}{2}}\,
\int_{\epsilon}^1\,\frac{dy}{y^\alpha}\,\sqrt{1-y^2}.
\end{equation}
The asymptotic behavior of $\langle{\rm
Tr}_{\kappa_{n,\beta}}\,|D|^{-\alpha}\rangle_{n,\beta}$ does not depend
in an essential way on the particular choice of $\epsilon$, as
long as we keep $\epsilon\neq0$.

Now we use the asymptotic formula (\ref{30}) in (\ref{28}) and
search for the value $\alpha_c$ for which, as $N\to\infty$ and
$\beta\to\beta_c$, $\langle{\rm
Tr}_{\kappa_{n,\beta}}\,|D|^{-\alpha}\rangle$ diverges (converges to
zero) if $\alpha<\alpha_c$ ($\alpha>\alpha_c$), with $\langle{\rm
Tr}_{\kappa_{n,\beta}}\,|D|^{-\alpha_c}\rangle$ finite and non-zero. 
This gives an upper bound for the dimension of the \lq\lq condensed'' 
manifold in the infinite phase ( $\beta\leq\beta_c$), which is
$\langle\delta\rangle< \alpha_c$. We obtain:

\begin{equation}\label{31} 
\langle{\rm Tr}_{\kappa_{n,\beta}}\,|D|^{-\alpha}
\rangle(\beta)\,\sim\,\lim_{N\to\infty}\,
\sum_{n=1}^{N}\, P(n,\beta)\,n^{1-\frac{\alpha}{2}}~.
\end{equation} 
In the finite phase ($\beta>\beta_c$) the sum
in (\ref{31}) converges for $\alpha\geq 0 $. We conclude 
that  $\alpha_c=0$
(i.e. $\langle\delta\rangle=0$) for $\beta>\beta_c$ , as expected.
From the behavior of $P(n,\beta)$ in the 
infinite phase it follows that the convergence  of
the sum in (\ref{31}) is dictated by the behavior of
$\Gamma_{n,\alpha}= n^{1-\frac{\alpha}{2}}$ in the
limit $n\sim N\to\infty$.  For
$\beta\leq\beta_c$ we get $\Gamma_{n,\alpha}\to\infty$ if
$\alpha<2$, and $\Gamma_{n,\alpha}\to0$ if $\alpha>2$. For
$\alpha=2$ it turns out that $\langle{\rm
Tr}_{\kappa_{n,\beta}}\,|D|^{-2}\rangle(\beta)\,\sim 1$.
Therefore, we obtain the upper bound 
$\langle\delta\rangle < 2$.

Notice that we do not have a definition of the dimension as 
an operator (as suggested by (\ref{19})), from which it will be 
possible to  compute its
average in the ensemble of all Hermitian matrices, like we have do
for $n$ in (\ref{11}). Instead, our guide is the operational
definition encapsulated in (\ref{28}). 
In order to go beyond the upper bound computed above we need to
find a suitable observable which reduces to the
classical dimension (as given by the Weyl formula) in the limit
$n\to\infty$. This eventually may lead to a numerical computation
of $\langle\delta\rangle$. Besides, one sees that the reason for the upper
bound $\langle\delta\rangle < 2 $ lies in the 
semi-circle law and
its leading $\sqrt{\lambda}$ behavior near the edge. It is known
from $2D$ models of discretized pure quantum gravity
\cite{Bre,Kan} that some special matrix polynomial potentials
$V(D)=\sum a_k {\rm Tr}D^{2k}$ may lead, in a suitable scaling
limit, to a behavior near the edges different from the square
root. Thus, a possible way to obtain a bound of higher dimension
would be to include higher polynomial interactions, or modify the
quadratic one in (\ref{4}) including more (internal) symmetries
besides the unitary one. These and other questions are under study,
and will be reported elsewhere.

To conclude, in this Letter we proposed a discrete model for quantum
gravity based on the spectral principle. The model is connected with 
the GUE of Hermitian matrices, and contains the mean number of points, 
$\langle n\rangle$,
and the dimension of the space-time, $\langle\delta\rangle$, as dynamical
observables. We have shown that the model has two phases: a finite
phase with a finite value of $\langle n\rangle$ and 
$\langle\delta\rangle=0$, and an
infinite phase with a diverging $\langle n\rangle$ and a finite 
$\langle\delta\rangle\ne0$.
The critical point was computed, $\beta_c=\pi/2$, as well as the
critical exponent of $\langle n\rangle$. Moreover, an upper bound 
for the order
parameter $\langle\delta\rangle$ was found, $\langle\delta\rangle < 2$.

We thank  S. Vaidya for
suggestions and valuable comments, A. P. Balachandran, D. Dalmazi,
J. C. A. Barata and A. Bertuola for helpful conversations. 
L.C.A. would like to
thank the Mathematical Physics Department of USP at S\~ao Paulo
for their kind hospitality. This work was partially supported by
FAPESP, grant 00/03277-3 (L.C.A.), and CNPq (P.T.S.).


\begin{references}
\bibitem{footnote}  It can be shown that the leading
singularity, for $\beta\leq \beta_c$, comes from the integral term
in (\ref{12}).
\end{references}
\end{document}